\documentclass[12pt,a4paper]{article}

\usepackage{cite}
\usepackage{amssymb}
\usepackage{graphicx}

\newcommand{\Angle}{\measuredangle}

\newcommand{\dd}{\mathrm{d}}
\newcommand{\GeV}{\ensuremath{\;\mathrm{GeV}}}
\newcommand{\Eep}{E_e{\!\!'}\,}
\newcommand{\pep}{p_e{\!\!'}\,}

\newcommand{\vareps}{\varepsilon}
\newcommand{\Li}{\mbox{Li}_2}

\begin{document}

%%% Enable double spacing if desired:
% \renewcommand{\baselinestretch}{2}\normalsize

%%%%%%%%%%%%%%%%%%%%%%%%%%%%%%%%%%%%%%%%%%%%%%%%%%%%%%%%%%%%%%%%%%%%%%%%
%%% Title Page starts here:

\begin{titlepage}

% We are supposed to give proper credit...
\renewcommand{\thefootnote}{\fnsymbol{footnote}}
\setcounter{footnote}{0}

%%% Preprint no.:
\begin{flushright}
  SI-98-32 \\
  hep-ph/9901258 \\
  January 1999
\end{flushright}

\vspace*{0.5cm}

\begin{center}
  {\Large\textbf{
    Deep Inelastic Scattering               \\[0.5ex]
    with a Tagged Photon:                   \\[0.5ex]
    QED Corrections for the $\Sigma$ Method \\[0.5ex]
  }}
\end{center}

% Slightly shortened title, if needed:
%\markboth{DIS with a Tagged Photon: QED Corrections for the $\Sigma$ Method}

\vspace*{0.5cm}

%%% Author(s) & Affiliation(s):

\begin{center}
  {\large Harald Anlauf\,%
  \footnote{Supported by Bundesministerium f\"ur Bildung, Wissenschaft,
            Forschung und Technologie (BMBF), Germany.}%
  \footnote{Email: \texttt{anlauf@hep.tu-darmstadt.de}}
  } \\[1ex]
  \textit{Fachbereich Physik, Siegen University, 57068 Siegen, Germany}
\end{center}

\vspace*{0.5cm}

% \abstract{
\begin{abstract}
  After a brief review of the kinematics of deep inelastic scattering (DIS)
  within the so-called $\Sigma$ method, we derive the necessary formulae for
  the treatment of QED radiative corrections to DIS originating from hard
  photon radiation.  The results are applied to a calculation of the
  corrections to DIS with a tagged photon with next-to-leading logarithmic
  accuracy under HERA conditions.  It turns out that the next-to-leading
  logarithmic corrections are quite important for the $\Sigma$ method.  We
  also discuss the dependence of the corrections on the longitudinal structure
  function of the proton, $F_L$, in the region of low $Q^2$ and moderate $x$.
\end{abstract}
%   \PACS{
%     {13.60.--r}{Photon and charged-lepton interactions with hadrons} \and
%     {13.60.Hb}{Total and inclusive cross sections} \and
%     {12.15.Lk} Electroweak radiative corrections}
%   }
% }

\end{titlepage}

%%% End Title Page
%%%%%%%%%%%%%%%%%%%%%%%%%%%%%%%%%%%%%%%%%%%%%%%%%%%%%%%%%%%%%%%%%%%%%%%%

\renewcommand{\thefootnote}{(\arabic{footnote})}
\setcounter{footnote}{0}

%%% No Macros are defined or redefined below.

%%%%%%%%%%%%%%%%%%%%%%%%%%%%%%%%%%%%%%%%%%%%%%%%%%%%%%%%%%%%%%%%%%%%%%%%
%%% Text begins here:

\section{Introduction}
\label{sec:Intro}

The determination of the structure functions of the proton, $F_2(x,Q^2)$ and
$F_L(x,Q^2)$, over a broad range of the kinematic variables belongs to the
most important tasks of the H1 and ZEUS experiments at the HERA $ep$ collider.
Especially the extension of these measurements to the range of small Bjorken
$x < 10^{-4}$ and $Q^2$ of a few GeV$^2$ is of particular interest, as it
provides a testing ground for our attempts to understand the details of the
dynamics of quarks and gluons inside the nucleon.

Whilst the structure function $F_2$ can be extracted quite easily from the
experimental data, it is more difficult to determine the longitudinal
structure function $F_L$.  A direct method that relies only on measured data
requires running the collider at different center-of-mass energies.  However,
besides impairing the high-energy program of the machine, running at reduced
beam energies also increases some systematic errors, (e.g., luminosity
uncertainties), in the experimental analysis.

These problems are circumvented by employing a method suggested by Krasny et
al.~\cite{KPS92} that utilizes radiative events.  This method takes advantage
of a photon detector (PD) in the very forward direction, as seen from the
incoming lepton (electron or positron) beam.  Such a device is part of the
luminosity monitoring system of both the H1 and ZEUS experiments.

The idea of this method is that emission of photons in a direction close to
the incoming lepton corresponds to a reduction of the effective beam energy.
This effective beam energy for each radiative event is determined from the
energy of the hard photon observed (tagged) in the PD.  Early analyses that
make use of these radiative events for a determination of $F_2$ were already
published in \cite{H1:rad,ZEUS96}.  No QED radiative corrections were taken
into account in these analyses.  The feasibility of a determination of $F_L$
was studied in \cite{FGMZ96}.

Recently, the H1 collaboration presented preliminary results of a refined
analysis with newer data \cite{H1:ISR-prelim}.  In this analysis, the authors
chose different methods of determination of the kinematic variables%
\footnote{For a discussion of the most common methods to determine kinematic
  variables and further references, see e.g.\ \cite{Wol97}.}
(the $e$-method, where the kinematic variables are obtained from a measurement
of the scattered lepton, and the $\Sigma$ method) in different $(x,Q^2)$ bins
in order to reduce the experimental systematic error.  However, since the
calculations of the QED radiative corrections to DIS with a tagged photon
\cite{AAKM:ll,AAKM:JETP,AAKM:nlo} did not cover the $\Sigma$ method, the
corrections were only applied to part of the data in~\cite{H1:ISR-prelim}.  It
is the purpose of the present work to extend these analytical calculations to
the $\Sigma$ method.

The $\Sigma$ method, as proposed by Bassler and Bernardi \cite{BB95}, tries to
combine the momenta of the outgoing lepton and hadrons judiciously in order to
reduce experimental systematic uncertainties on the determination of the
kinematic variables especially in the kinematic region of low $Q^2$ where
other methods are limited by e.g., detector resolution or energy calibration.

With the help of the quantity%
\footnote{Note that in this paper we take the positive $z$-axis along the
  initial lepton direction, unlike \cite{BB95} who chose the direction of the
  incoming proton beam.}
\begin{equation} \label{eq:sigma-BB}
  \Sigma_h \, \equiv \, \sum_h \left( E_h + p_{z,h} \right) ,
\end{equation}
where the sum runs over the detected hadrons, and $E_h$ and $p_{z,h}$ are the
energy and $z$-component of the respective particle, the kinematic variables
$x_\Sigma$, $y_\Sigma$ and $Q^2_\Sigma$ are defined via
\begin{equation} \label{eq:xyQ2-BB}
  y_\Sigma = \frac{\Sigma_h}{\Sigma_h + \Eep (1 + \cos\theta)}
  \, , \quad
  Q^2_\Sigma = \frac{\Eep^2 \sin^2 \theta}{1 - y_\Sigma}
  \, , \quad
  x_\Sigma = \frac{Q^2_\Sigma}{y_\Sigma S}
  \, .
\end{equation}
Here $S = 4 E_e E_p$, where $E_e$ and $E_p$ are the beam energies of the
lepton and proton beam, respectively, and $\Eep$ and $\theta$ are the energy
and scattering angle of the outgoing lepton, measured with respect to the
direction of the initial lepton.

One of the known advantages of the $\Sigma$ method is its insensitivity of the
determination of $y_\Sigma$ and $Q^2_\Sigma$ to undetected emission of a hard
photon collinear to the incoming lepton (initial state radiation, ISR).

The $\Sigma$ method has already been used in several analyses of the H1
collaboration.  However, the author is not aware of any publications on an
analytical (i.e., non-Monte Carlo) treatment of QED radiative corrections to
deep inelastic scattering (DIS) using the $\Sigma$ kinematic variables%
\footnote{As the tagged photon cross section represents a radiative correction
  to the DIS cross section \cite{AAKM:ll}, one can in principle calculate the
  QED corrections to the former for any choice of kinematic variables with the
  help of a Monte Carlo event generator for DIS that properly implements the
  necessary higher order QED corrections to DIS.  However, no generator exists
  for the calculation of QED corrections beyond leading logarithms, and the
  leading log generator KRONOS \cite{KRONOS} uses approximations for photon
  emission in the very forward direction that make it useless for the present
  task.}
beyond the collinear (leading log) approximation
\cite{HECTOR,BBCK96}.  The presumable reason is that the $\Sigma$ method has
only been introduced long after the start of data taking at HERA.  Therefore,
this paper starts with a brief introduction to the kinematics of radiative DIS
in the $\Sigma$ method.  Section~\ref{sec:tagged} extends the considerations
to the case of DIS with an exclusive tagged photon, but specialized to the
conditions at HERA, and provides the relevant formulae to calculate the
radiative corrections to the tagged photon cross section, based on the results
of \cite{AAKM:nlo}.  Some results for HERA experimental conditions are
presented in section~\ref{sec:results}, and section~\ref{sec:sum} contains our
conclusion.  Finally, the appendices collect several technical details.

%%%%%%%%%%%%%%%%%%%%%%%%%%%%%%%%%%%%%%%%%%%%%%%%%%%%%%%%%%%%%%%%%%%%%%%%

\section{Kinematics in the $\Sigma$ Method}
\label{sec:kinem}

This section is devoted to a basic review of the $\Sigma$ method.  Here we
shall prepare an appropriate framework for the treatment of radiative
corrections to radiative deep inelastic scattering,
\begin{equation} \label{eq:kin1}
  e(p) + P(P) \; \to \; e(p') + X(P') + \gamma(k) \, ,
\end{equation}
i.e., DIS with single hard photon emission.  The extension to the process with
an additional tagged photon in the forward direction is straightforward and
will be performed in the next section.

Let us begin by stating our conventions for the kinematics in the HERA lab
frame that are used throughout this paper.  We shall take the orientation of
the coordinate frame such that the positive $z$-axis points in the direction
of the incoming lepton beam, and the momentum of the scattered lepton lies in
the $x$-$z$-plane:
\begin{eqnarray} \label{eq:HERA-kin}
  P  & = & (E_p,0,0,-p_p) \, , \nonumber \\
  p  & = & (E_e,0,0,p_e)  \, , \nonumber \\
  p' & = & (\Eep,\pep \sin\theta,0,\pep \cos\theta) \, , \nonumber \\
  k  & = &
  E_\gamma (1, \sin\vartheta \cos\varphi, \sin\vartheta \sin\varphi, \cos\vartheta)
  \, .
\end{eqnarray}
As the beam energies $E_p, E_e$, as well as the energy of the scattered
lepton, $\Eep$, are always large compared to the proton mass, $M$, and the
electron mass, $m$, we shall take $p_p = E_p$, $p_e = E_e$, and $\pep = \Eep$,
wherever possible.

Since we assume $E_p \gg M$, we may replace the definition of the variable
$\Sigma_h$ in (\ref{eq:sigma-BB}) by
\begin{equation} \label{eq:sigma-ha}
  \Sigma_h \, := \, \frac{P \cdot (P' - P)}{E_p} \, .
\end{equation}
This allows us to similarly reexpress the definitions (\ref{eq:xyQ2-BB}) of
the kinematic variables $x_\Sigma$, $y_\Sigma$, and $Q_\Sigma^2$ through
scalar products of four-momenta via
\begin{eqnarray} \label{def:y}
  y_\Sigma & = & \frac{P \cdot (P' - P)}{P \cdot (P' - P + p')} \, ,
  \\
  \label{def:Q2}
  Q_\Sigma^2 (1 - y_\Sigma) & = & \frac{4 (p \cdot p')(P \cdot p') }{S}
  \, , \\
  \label{def:x}
  x_\Sigma & = & \frac{Q^2_\Sigma}{y_\Sigma S}
  \, , \\
\noalign{\hbox{where}}
  S & = & 2 P \cdot p
  \, . \nonumber
\end{eqnarray}
One important thing to note here is the nonlinear dependence of the kinematic
variables $y_\Sigma$ and $Q^2_\Sigma$ on the energy and scattering angle of
the outgoing lepton, while it is linear in the electron-only method
($e$-method).

As we are dealing with the kinematics of a process with real photon emission,
it is convenient to define an invariant quantity $\kappa$, that represents the
energy of the outgoing photon in units of the energy of the incoming lepton,
as measured in the rest frame of the incoming proton,
\begin{equation} \label{def:kappa}
  \kappa \; := \; \frac{P \cdot k}{P \cdot p} \, .
\end{equation}
Energy and momentum conservation of the process (\ref{eq:kin1}) obviously
requires $0 \leq \kappa < 1$.  In the special case of emission of the photon
collinear to the incoming lepton, $\kappa$ also represents the energy fraction
of the initial lepton taken by the photon in the HERA lab frame.

Using momentum conservation (\ref{eq:kin1}) and relations (\ref{def:y}) and
(\ref{def:kappa}), it is easy to see that
\begin{equation} \label{eq:Pdotq}
  P \cdot (P' - P)
  \; = \; (1-\kappa) y_\Sigma \, P \cdot p
  \; = \; (1-\kappa) y_\Sigma \cdot \frac{1}{2} S
  \, .
\end{equation}

To proceed, let us express the remaining scalar products of external momenta
in terms of invariant kinematic variables and other measured quantities.  We
find
\begin{eqnarray} \label{eq:scprod1}
  P \cdot p' & = & \frac{1 - y_\Sigma}{y_\Sigma} \, P \cdot (P' - P)
        \; = \; (1-\kappa) (1 - y_\Sigma) \cdot \frac{1}{2} S \, ,
        \nonumber \\
  p \cdot p' & = &
        \frac{x_\Sigma \, y_\Sigma^2 \, S^2}{4 P \cdot (P' - P)}
        \; = \; \frac{x_\Sigma \, y_\Sigma}{1-\kappa} \cdot \frac{1}{2} S \, .
\end{eqnarray}
Furthermore, from the photon's energy $E_\gamma =: x_\gamma E_e$ and angles
$\vartheta$ and $\vartheta'$ with respect to the incoming and final lepton,
respectively, we obtain
\begin{eqnarray} \label{eq:scprod2}
  p  \cdot k & = & x_\gamma E_e^2 (1 - \cos\vartheta)
  \, , \nonumber \\
  p' \cdot k & = & x_\gamma E_e \Eep (1 - \cos\vartheta')
  \, ,
\end{eqnarray}
where the angle $\vartheta'$ is calculated from (\ref{eq:HERA-kin}),
\begin{equation} \label{eq:c_f}
  \cos\vartheta' =
  \cos\theta \cos\vartheta + \sin\theta \sin\vartheta \cos\varphi \, .
\end{equation}

Next, we rewrite the left-hand sides of the relations (\ref{eq:scprod1}) in
the HERA lab frame as
\begin{eqnarray}
  P \cdot p' & = & E_p \Eep (1 + \cos\theta)
  \, , \nonumber \\
  p \cdot p' & = & E_e \Eep (1 - \cos\theta)
  \, ,
\end{eqnarray}
to derive explicit expressions for the energy and scattering angle of the
outgoing lepton in the HERA frame,
\begin{eqnarray} \label{eq:Eep}
  \Eep & = &
  (1-\kappa) (1 - y_\Sigma) E_e + \frac{x_\Sigma y_\Sigma}{1-\kappa} E_p
  \, , \nonumber \\
  \cos\theta & = &
  \frac{(1-\kappa)^2 (1 - y_\Sigma) E_e - x_\Sigma y_\Sigma E_p}
       {(1-\kappa)^2 (1 - y_\Sigma) E_e + x_\Sigma y_\Sigma E_p}
  \, ,
\end{eqnarray}
that may further be used to eliminate $\Eep$ in the second equation of
(\ref{eq:scprod2}), or to trade the azimuthal angle $\varphi$ in favor of the
angle $\vartheta'$.

It is now straightforward to obtain the expressions for the momentum transfer
to the hadronic system
\begin{equation} \label{eq:Qh2}
  Q_h^2 \; \equiv \; - (P-P')^2
        \; = \; - (p-p'-k)^2
        \; = \; \frac{x_\Sigma y_\Sigma S}{1-\kappa} + 2 k \cdot (p-p')
        \, ,
\end{equation}
the hadronic scaling variable
\begin{equation} \label{eq:xh}
  x_h \equiv \frac{Q_h^2}{2 P \cdot (p-p'-k)}
     \; = \; \frac{Q_h^2}{(1-\kappa) y_\Sigma S}
     \, ,
\end{equation}
and the invariant mass of the hadronic system,
\begin{eqnarray} \label{eq:W2}
  W^2 & = & (P')^2 \; = \; (P + p - p' - k)^2
  \\
      & = &
        M^2
        + y_\Sigma \left[ 1 - \kappa - \frac{x_\Sigma}{(1 - \kappa)} \right] S
        - 2 k \cdot (p-p')
        \; = \; M^2 + \frac{1 - x_h}{x_h} Q_h^2
  \nonumber \, .
\end{eqnarray}
In the last equation we explicitly retained the proton mass.

The kinematic limit for the phase space of the radiated photon may be derived
by requiring that the invariant mass of the hadronic system be larger than the
threshold for pion production,
\begin{equation}
  W^2 \geq \bar{M}^2
  \, , \qquad \mbox{where} \quad
  \bar{M} = M + m_\pi \, .
\end{equation}
The actual upper limit on the photon energy as a function of the emission
angle, $E_\gamma^\mathrm{max} = E_\gamma^\mathrm{max}(\theta; \vartheta,
\varphi) = E_\gamma^\mathrm{max}(\theta; \vartheta, \vartheta') $ is obtained
in general by solving a quadratic or cubic equation; for details see
appendices~\ref{app:kin-limit1} and~\ref{app:kin-limit2}.

Before we conclude this section, we shall give the relations between the
kinematic variables in the $\Sigma$ method and in the $e$-only method.
{}From the scalar products (\ref{eq:scprod1}), we find:
\begin{eqnarray}
  1-y_e \; = \; (1-\kappa)(1-y_\Sigma)
  & \Longrightarrow &
  y_e \; = \; y_\Sigma + \kappa(1 - y_\Sigma)
  \, , \nonumber \\
  x_e y_e \; = \; \frac{x_\Sigma y_\Sigma}{1-\kappa}
  & \Longrightarrow &
  x_e \; = \;
  \frac{x_\Sigma y_\Sigma}{(1 - y_\Sigma) [y_\Sigma + \kappa(1 - y_\Sigma)]}
  \, .
\end{eqnarray}
The Jacobian between these two sets is
\begin{equation}
  \mathcal{J} \; \equiv \;
  \det \left( \frac{\partial(x_e,y_e)}{\partial(x_\Sigma,y_\Sigma)} \right)
  \; = \;
  \frac{y_\Sigma}{y_\Sigma + \kappa(1 - y_\Sigma)}
  \; = \;
  \frac{y_\Sigma}{y_e} \, .
\end{equation}

The reader may easily verify that the above formulae are consistent with the
collinear limits discussed in \cite{HECTOR}.

%%%%%%%%%%%%%%%%%%%%%%%%%%%%%%%%%%%%%%%%%%%%%%%%%%%%%%%%%%%%%%%%%%%%%%%%

\section{DIS with a Tagged Photon}
\label{sec:tagged}

After having discussed the kinematics in the $\Sigma$ method, let us now turn
to our primary aim, the description of radiative corrections to neutral
current deep inelastic scattering with an exclusive tagged photon for the HERA
collider.

%%%

\subsection{Kinematics and Lowest Order Cross Section}

As already explained in the introduction and described in more detail in
\cite{KPS92} (and references cited therein), the experimental detection of
photons emitted in the very forward direction is actually possible at the HERA
collider due to the presence of photon detectors (PD) that are part of the
luminosity monitoring systems of the H1 and ZEUS experiments.  These PD's
cover an angular range very close to the direction of the incoming lepton
beam, $(\vartheta_\gamma \equiv \Angle(\vec{p},\vec{k}) \leq \vartheta_0
\approx 5\cdot 10^{-4}$~rad).

Thus, the process under consideration corresponds to the reaction
\begin{equation} \label{eq:tag-kin}
  e(p) + p(P) \; \to \;
  e(p') + \gamma(k) + X(P') + \left[ \gamma(k_2) \right] \, ,
\end{equation}
where $\gamma(k)$ denotes the collinearly emitted, exclusively
measured (tagged) photon, and $\left[ \gamma(k_2) \right]$ indicates
an additional (i.e., second) photon in the case of the radiative
correction to the lowest order process.

A set of kinematic variables that is adapted to take into account tagged
collinear radiation is given by the \textit{shifted} Bjorken variables
\cite{KPS92,AAKM:ll,AAKM:nlo}.%
\footnote{Here we shall use the notation $\hat{x}$ etc.\ of \cite{AAKM:nlo} to
  avoid confusion between shifted Bjorken variables and the usual ones.
  Although $\hat{x}$ etc.\ are determined in the $\Sigma$ method, we drop the
  index $\Sigma$ in order to not overload the notation.}
Expressed in terms of the measured quantities, $\Sigma_h$ (see
eq.~\ref{eq:sigma-ha}), the energy $\Eep$ and angle $\theta$ of the scattered
lepton, and the energy of the tagged photon, they read
\begin{equation} \label{eq:kin-vars-measured}
  \hat{y} = \frac{\Sigma_h}{\Sigma_h + \Eep (1+\cos\theta)} \, ,
  \quad
  \hat{Q}^2 = \frac{\Eep{}^2 \sin^2\theta}{1 - \hat{y}} \, ,
  \quad
  \hat{x} = \frac{\hat{Q}^2}{\hat{y} zS } \, .
\end{equation}
Here we denoted by $z$ the energy fraction of the lepton after initial state
radiation of a collinear photon,
\begin{equation} \label{def:z}
  z = \frac{2P \cdot (p - k)}{S} = \frac{E_e - E_\gamma}{E_e} \, ,
\end{equation}
where $E_\gamma$ represents the energy deposited in the forward PD.

The above definition (\ref{eq:kin-vars-measured}) corresponds to
eqs.~(\ref{def:y})--(\ref{def:x}), but with an effectively reduced
center-of-mass energy, $zS$.  It is obvious that in the case of the $\Sigma$
method only $\hat{x}$ is affected by collinear initial state radiation, as was
already found in \cite{BB95}.

In analogy to the case of ordinary DIS, we may rewrite the kinematic variables
in a Lorentz invariant fashion:
\begin{equation} \label{eq:kin-vars}
  \hat{y}   = \frac{P \cdot (P' - P)}{P \cdot (P' - P + p')} \, ,
  \quad
  \hat{Q}^2 = \frac{4 (zp \cdot p')(P \cdot p') }{(1 - \hat{y}) zS} \, ,
  \quad
  \hat{x}   = \frac{\hat{Q}^2}{\hat{y} zS} \, .
\end{equation}

The Born cross section, integrated over the solid angle of the photon detector
($0 \leq \vartheta_\gamma \leq \vartheta_0, \; \vartheta_0 \ll \theta$) takes
a factorized form (see also \cite{KPS92,BKR97,AAKM:ll,AAKM:nlo}):
\begin{equation} \label{eq:Born}
  \frac{1}{\hat{y}} \,
  \frac{\dd^3\sigma_\mathrm{Born}}{\dd\hat{x}\,\dd\hat{y}\,\dd z} =
  \frac{\alpha}{2\pi} \, P(z,L_0) \, \tilde\Sigma \, ,
\end{equation}
where
\begin{eqnarray} \label{eq:Sigma-tilde}
  \tilde\Sigma
  & \equiv & \tilde\Sigma(\hat{x},\hat{y},\hat{Q}^2)
  \nonumber \\
  & = & \frac{2\pi\alpha^2(-\hat{Q}^2)}{\hat{Q}^2\hat{x}\hat{y}^2}
  \left[ 2(1-\hat{y}) - 2\hat{x}^2\hat{y}^2\frac{M^2}{\hat{Q}^2}
    + \left( 1+4\hat{x}^2\frac{M^2}{\hat{Q}^2}\right) \frac{\hat{y}^2}{1+R}
  \right]
  F_2(\hat{x},\hat{Q}^2) ,
  \nonumber
\end{eqnarray}
with
\begin{eqnarray}
  P(z,L_0) & = & \frac{1+z^2}{1-z} L_0 - \frac{2z}{1-z} \, , \qquad
  L_0 = \ln\left(\frac{E_e^2\vartheta_0^2}{m^2}\right)     , \nonumber \\
  \hat{Q}^2 & = & \hat{x}\hat{y} zS \, , \qquad
  \alpha(-\hat{Q}^2) = \frac{\alpha}{1-\Pi(-\hat{Q}^2)} \, , \nonumber \\
  \label{eq:R}
  R = R(\hat{x},\hat{Q}^2) & = &
  \left( 1+4\hat{x}^2\frac{M^2}{\hat{Q}^2} \right)
  \frac{F_2(\hat{x},\hat{Q}^2)}{2\hat{x}F_1(\hat{x},\hat{Q}^2)} \, - 1
  \, .
\end{eqnarray}
The quantities $F_2$ and $F_1$ denote the proton structure functions.  Note
that we explicitly include the correction from vacuum polarization
$\Pi(-\hat{Q}^2)$ in the virtual photon propagator, and that we neglect the
contributions from Z-boson exchange and $\gamma$-Z interference, because we
are interested mostly in the kinematic region of small momentum transfer
$\hat{Q}^2$.

%%%

The cross section (\ref{eq:Born}) describes the process (\ref{eq:tag-kin}) to
lowest order in perturbation theory.  The radiative corrections to this cross
section are composed of contributions by corrections due to virtual photon
exchange, soft photon emission, and emission of a second hard photon, with one
of the hard photons being tagged in the PD.  Because of its coarse
granularity, we shall assume that the PD cannot measure photons individually
but only their total energy when two hard photons simultaneously hit the PD in
different locations.

%%%

\subsection{Virtual and Soft Corrections}

The virtual and soft corrections to the lowest order cross section are simply
obtained from the calculation for a measurement using the $e$-method by
substitution.  For the sake of completeness, we quote the result from
\cite{AAKM:nlo}:
\begin{equation} \label{eq:V+S}
  \frac{1}{\hat{y}} \,
  \frac{\dd^3\sigma_\mathrm{V+S}}{\dd\hat{x}\,\dd\hat{y}\,\dd z} =
  \frac{\alpha^2}{4\pi^2} \left[ P(z,L_0) \tilde{\rho} - T \right]
  \tilde\Sigma \, ,
\end{equation}
with
\begin{eqnarray} \label{eq:rho-tilde}
  \tilde{\rho} & = &
  2(L_Q - 1)\ln\frac{\Delta^2}{Y}
  + 3L_Q + 3\ln z - \ln^2Y - \frac{\pi^2}{3} - \frac{9}{2}
  + 2\Li \left( \frac{1+c}{2} \right) ,
  \nonumber \\
  T & = &
  \frac{1+z^2}{1-z}(A\ln z + B)
  - \frac{4z}{1-z}L_Q\ln z
  - \frac{2-(1-z)^2}{2(1-z)} L_0 + {\mathcal O}({\mathrm{const}}) \, ,
  \nonumber \\
  A & = & {} -L_0^2 + 2L_0L_Q - 2L_0\ln(1-z) \, ,
  \quad
  B   =   \left[ \ln^2 z - 2 \Li (1-z)\right] L_0 \, ,
  \nonumber \\
  L_Q & = & \ln\frac{\hat{Q}^2}{z m^2} \, , \qquad
  \Li (t) = -\int\limits_{0}^{t}\frac{\dd u}{u}\ln(1-u) \, .
\end{eqnarray}
Here $\Delta$ denotes the infrared cutoff for the emission of a soft photon in
addition to the hard one, $\Delta = E_{\gamma2}^\mathrm{max}/E_e$.
Furthermore,
\begin{equation} \label{eq:Y-c-elastic}
  Y \equiv \frac{\Eep}{E_e}
  = z(1 - \hat{y}) + \hat{x}\hat{y} \, \frac{E_p}{E_e}
  \quad \mbox{and} \quad
  c \equiv \cos\theta =
  \frac{z(1 - \hat{y}) E_e - \hat{x}\hat{y} E_p}
       {z(1 - \hat{y}) E_e + \hat{x}\hat{y} E_p}
\end{equation}
follow from the formulae of the previous section with the replacement $p \to
zp$, as the energy loss due to the tagged collinear photon is known and
already taken into account in the determination of the kinematic variables.

It should be noted that in (\ref{eq:V+S})--(\ref{eq:rho-tilde}) and also
further below we retain only terms with double or single large logarithms of
the small electron mass $m$, i.e., terms of order $\alpha^2 L^2$ and $\alpha^2
L$, with $L$ being one of $L_0$ or $L_Q$.  As the lowest order cross section
(\ref{eq:Born}) is of order $\alpha L$ relative to the DIS cross section, we
shall denote the terms of order $\alpha^2 L^2$ as leading (LL) and those or
order $\alpha^2 L$ as next-to-leading logarithmic (NLL) ones.

%%%

\subsection{Double Hard Bremsstrahlung}

Besides the soft and virtual corrections to the lowest order process, we have
to consider also the corrections from hard bremsstrahlung, which in the
present case corresponds to double hard bremsstrahlung.

In the calculation of the contributions from the emission of two hard photons,
it is convenient to decompose the phase space into same three regions
discussed in \cite{AAKM:nlo}: \textit{i)} both hard photons hit the forward
photon detector, i.e., both are emitted within a narrow cone around the lepton
beam $(\vartheta_{1,2} \leq \vartheta_0, \, \vartheta_0 \ll 1)$; \textit{ii)}
one photon is tagged in the PD, while the other is collinear to the outgoing
lepton $(\vartheta_2' \equiv \Angle(\vec{k}_2,\vec{p}{\,'}) \leq
\vartheta'_0)$; and finally \textit{iii)} the second photon is emitted at
large angles (i.e., outside the defined narrow cones) with respect to both
incoming and outgoing lepton momenta.  The last kinematic domain is denoted as
the semi-collinear one.  For the sake of simplicity, we shall always assume
below that $\vartheta'_0 \ll 1$.

The contribution from the kinematic region \textit{i)} (both hard photons
being tagged, and only the sum of their energies measured), has the form
\cite{AAKM:nlo}:
\begin{eqnarray} \label{eq:sig-i}
  \frac{1}{\hat{y}} \,
  \frac{\dd^3 \sigma^{\gamma\gamma}_i}{\dd\hat{x}\,\dd\hat{y}\, \dd z}
  & = &
  \frac{\alpha^2}{8\pi^2} L_0
  \Biggl[ L_0 \left( P^{(2)}_{\Theta}(z) +
          2\frac{1+z^2}{1-z}\left(\ln z-\frac{3}{2}-2\ln\Delta\right)\right)
  \nonumber \\
  &&
  {} + 6(1-z) + \left(\frac{4}{1-z}-1-z\right)\ln^2z - 4\frac{(1+z)^2}{1-z}
  \ln\frac{1-z}{\Delta}\Biggr] \tilde\Sigma
  \nonumber\\ &&
  {} + {\mathcal O}({\mathrm{const}}) \, ,
\end{eqnarray}
with
\[
  P^{(2)}_{\Theta}(z) =
  2 \left[ \frac{1+z^2}{1-z} \left( 2\ln(1-z)-\ln z +\frac{3}{2} \right)
         + \frac{1}{2}(1+z)\ln z-1+z \right] .
\]

%%%

The contributions to the kinematic regions \textit{ii)} and \textit{iii)} in
the present case are slightly more complicated than for the purely leptonic
measurement described in \cite{AAKM:nlo}.  Before discussing the contributions
to the cross section from radiation almost collinear to the final state
lepton, we shall therefore extend our treatment of the kinematics now and
exhibit the necessary changes to the notation introduced in section~2.

To this end, let us recall the introduction (\ref{def:kappa}) of the variable
$\kappa$.  Again we define
\begin{equation} \label{def:kappa2}
  \kappa \; := \; \frac{P \cdot k_2}{P \cdot p}
  \; = \; x_2 \, \frac{1 + \cos\vartheta}{2}
  \, .
\end{equation}
Here $x_2$ is the energy of the second (i.e., non-collinear) photon that is
not tagged in the PD, in units of the initial lepton energy, and $\vartheta$
is its emission angle with respect to the incoming lepton.  By similar
reasoning as in the previous section, one can easily see that we presently
have $0 \leq \kappa < z$.

On the other hand, the tagging of the collinear hard photon with energy
$E_\gamma$ in the PD corresponds to a reduction of the effective initial
lepton energy by a factor of $z$ (see eq.~\ref{def:z}).  The corresponding
relations between measured quantities and invariant variables are simply
obtained from eqs.~(\ref{eq:Pdotq})--(\ref{eq:W2}) via the simultaneous
substitutions $E_e \to z E_e, \, S \to zS, \, \kappa \to \kappa/z$.
Therefore, e.g., eqs.~(\ref{eq:Eep}) now read
\begin{eqnarray} \label{eq:Eep-shifted}
  \Eep & = &
  (z-\kappa) (1 - \hat{y}) E_e + \frac{\hat{x} \hat{y} z}{z-\kappa} E_p
  \, , \nonumber \\
  \cos\theta & = &
  \frac{(z-\kappa)^2 (1 - \hat{y}) E_e - \hat{x} \hat{y} z E_p}
       {(z-\kappa)^2 (1 - \hat{y}) E_e + \hat{x} \hat{y} z E_p}
  \, .
\end{eqnarray}
In addition we define
\begin{eqnarray} \label{eq:s-t}
  \hat{Q}_l^2 & \equiv & 2 zp \cdot p' =
  \frac{z \hat{Q}^2}{z - \kappa} =
  \frac{\hat{x} \hat{y} z^2 S}{z - \kappa}
  \, , \nonumber \\
  \hat{t} & \equiv & -2z p \cdot k_2 =
  2 x_2 z E_e^2 (1 - \cos\vartheta)
  \, , \nonumber \\
  \hat{s} & \equiv &  2 p' \cdot k_2 =
  2 x_2 E_e \Eep (1 - \cos\vartheta')
  \, . \nonumber
\end{eqnarray}
The momentum transfer to the hadronic system and the true hadronic scaling
variable are
\begin{eqnarray} \label{eq:Qh2-shifted}
  Q_h^2 & = & - (zp-p'-k_2)^2 = \hat{Q}_l^2 - \hat{t} - \hat{s}
  \, , \nonumber \\
  x_h   & = & \frac{Q_h^2}{2 P \cdot (zp-p'-k_2)} =
  \frac{Q_h^2}{(z-\kappa) \hat{y} S}
  \, .
\end{eqnarray}
The kinematic limit $x_2^\mathrm{max}(\vartheta,\vartheta')$ for the emission
of the second hard photon is again obtained by requiring that the hadronic
mass
\begin{equation} \label{eq:W2'}
  W^2 = (P + zp - p' - k_2)^2 = M^2 + \frac{1 - x_h}{x_h} Q_h^2
\end{equation}
be larger than the inelastic threshold, $\bar{M}^2$.  More details can be
found in appendix~\ref{app:kin-limit2}.

The contribution to the radiative cross section from the semi-collinear region
\textit{iii)} is obtained from the corresponding expression in the case of the
$e$-method (see eqs.~3.7f of ref.~\cite{AAKM:nlo}),
\begin{equation} \label{eq:sig-iii}
  \frac{1}{\hat{y}} \,
  \frac{\dd^3 \sigma^{\gamma\gamma}_{iii}}{\dd\hat{x}\,\dd\hat{y}\, \dd z}
  =
  \frac{\alpha^2}{\pi^2} P(z,L_0)
  \int\frac{\dd^3 k_2}{|\vec{k}_2|}
  \, \frac{\alpha^2(Q_h^2)}{Q_h^4}
  \, I^{\gamma}(zp,p',k_2)
  \, ,
\end{equation}
with
\begin{eqnarray} \label{eq:Igamma}
  I^{\gamma}
  & = &
  - \frac{1}{\hat{s}\hat{t}} \,
  \Biggl\{
  G F_1(x_{h},Q_{h}^2) +
  \Biggl[
  x_h \left[ z^2 + (z - \kappa)^2 (1-\hat{y})^2 \right] S^2
  - \frac{x_h M^2}{Q_h^2} \, G
  \nonumber \\
  &&
  \qquad \quad
  {} + \left[
    (z - \kappa) (1-\hat{y}) (\hat{Q}_l^2 - \hat{s})
    - z (\hat{Q}_l^2 - \hat{t})
  \right] S
  \Biggr]
  F_2(x_h,Q_h^2)
  \Biggr\} \, ,
  \nonumber \\
  G & = & Q_h^4 - 2 \hat{s}\hat{t} + \hat{Q}_l^4 \, ,
\end{eqnarray}
and $\hat{s}, \hat{t}, \hat{Q}_l^2, Q_h^2,$ and $x_h$ as defined above.  The
angular part of the $k_2$-integration is clearly restricted
%$\Theta$-function indicates the restriction of this contribution
to the kinematic region \textit{iii)}, i.e., the full solid angle with the
exception of the separately treated cones around the incoming and outgoing
lepton.

Let us finally turn to the discussion of the kinematic region \textit{ii)}.
As was already discussed in \cite{AAKM:nlo}, we expect that the contribution
of this region to the observed cross section will depend on the experimental
event selection, i.e., on the method of measurement of the scattered
particles.  We shall focus on the very same two possibilities.  The first one
is denoted as an \textit{exclusive} (or bare) event selection, as only the
scattered lepton is measured; the hard photon that is emitted almost
collinearly (i.e., within a small cone with opening angle $2\vartheta_0'$
around the momentum of the outgoing lepton) remains undetected or is not taken
into account in the determination of the kinematic variables.  The second,
more realistic case (from an experimental point of view) is a
\textit{calorimetric} event selection, when only the sum of the energies of
the outgoing lepton and photon is actually measured if the photon momentum
lies inside a small cone with opening angle $2\vartheta_0^{'}$ along the
direction of the final lepton.

First, in the case of an exclusive event selection, when only the scattered
lepton is detected, we obtain for $\vartheta_0' \ll 1$ (similarly to eq.~3.3
of \cite{AAKM:nlo})%
\footnote{We have of course kept the small but finite electron mass in the
  kinematic region of collinear radiation wherever necessary.  See
  e.g.~\cite{BBvdM91} for a very readable presentation.}
\begin{eqnarray} \label{eq:sig-ii-excl}
  \frac{1}{\hat{y}} \,
  \frac{\dd^3 \sigma^{\gamma\gamma}_{ii,\mathrm{excl}}}{\dd\hat{x}\,\dd\hat{y}\, \dd z}
  & = &
  \frac{\alpha^2}{4\pi^2} P(z,L_0)
  \int\limits_{\zeta_0}^{\zeta_\mathrm{max}}
  \frac{\dd \zeta}{\zeta^2}
  \left[ \frac{1+\zeta^2}{1-\zeta} \left(\widetilde L-1\right) + (1-\zeta) \right]
  \tilde\Sigma_f \, ,
  \nonumber \\
  \tilde\Sigma_f & \equiv & \tilde\Sigma(x_f,y_f,Q_f^2) \, ,
\end{eqnarray}
where
\begin{eqnarray}
  \widetilde L & = &
  L_0' + 2 \ln \frac{Y(\zeta)}{Y(1)} \, , \quad
  L_0' \equiv \ln \frac{E_e^2 \vartheta_0'{\!}^2}{m^2} + 2 \ln Y(1) \, ,
  \nonumber \\
  Y(\zeta) & = &
  \frac{\zeta^2 z(1-\hat{y}) +
    \left[ 1-\hat{y}(1-\zeta) \right]^2 \hat{x}\hat{y} \vareps}
  {\zeta \left[ 1-\hat{y}(1-\zeta) \right]}
  \nonumber \\
  x_f & = & \frac{\left[ 1-\hat{y}(1-\zeta) \right]^2}{\zeta^3} \, \hat{x}
  \, , \quad
  y_f = \frac{\zeta\hat{y}}{ 1-\hat{y}(1-\zeta)} \, ,
  \nonumber \\
  Q_f^2 & = & \frac{1-\hat{y}(1-\zeta)}{\zeta^2} \, \hat{Q}^2 \, ,
  \quad \vareps = \frac{E_p}{E_e} \, ,
  \nonumber \\
  \zeta_\mathrm{max} & = &
  1 - \frac{\Delta}{Y(1)} =
  1 - \frac{\Delta}{z(1-\hat{y}) + \hat{x}\hat{y} \vareps} \, ,
\end{eqnarray}
and $\zeta_0$ is the real solution in the interval $(0,1)$ of the cubic
equation
\begin{equation}
  \zeta_0^3 - \hat{x} \left[ 1 - \hat{y} (1-\zeta_0) \right]^2
  - \frac{\bar\Delta_m}{\hat{y} z}
  \left[ 1 - \hat{y} (1-\zeta_0) \right] \zeta_0^2 = 0 \, ,
\end{equation}
with $\bar\Delta_m = \left(\bar{M}^2 - M^2 \right)/S$, see also
appendix~\ref{app:kin-limit2}.  We note in passing that the form of the
leading logarithmic piece of (\ref{eq:sig-ii-excl}) agrees with the
corresponding terms of the calculation performed in \cite{HECTOR}.

Last, in the case of a calorimetric event selection, where only the sum of the
energies of the outgoing lepton and photon is measured if the photon momentum
lies inside the small cone with opening angle $2\vartheta_0^{'}$ along the
direction of the final lepton, the corresponding contribution reads
\begin{eqnarray} \label{eq:sig-ii-cal}
  \frac{1}{\hat{y}} \,
  \frac{\dd^3 \sigma^{\gamma\gamma}_{ii,\mathrm{cal}}}{\dd\hat{x}\,\dd\hat{y}\, \dd z}
  & = &
  \frac{\alpha^2}{4\pi^2} P(z,L_0)
  \int\limits_0^{\zeta_\mathrm{max}}
  \!\! \dd \zeta
  \left[ \frac{1+\zeta^2}{1-\zeta}
    \left(L_0' - 1 + 2\ln\zeta \right) + (1-\zeta) \right]
  \tilde\Sigma
  \nonumber \\
  & = &
  \frac{\alpha^2}{4\pi^2} P(z,L_0)
  \left[ \left( L_0' - 1 \right)
    \left(2 \ln \frac{Y(1)}{\Delta} - \frac{3}{2} \right)
    + 3 - \frac{2\pi^2}{3} \right]
  \tilde\Sigma \, . \qquad
\end{eqnarray}

The total contribution from QED radiative corrections is finally found by
adding up (\ref{eq:V+S}), (\ref{eq:sig-i}), (\ref{eq:sig-iii}), and, depending
on the chosen event selection, (\ref{eq:sig-ii-excl}) or
(\ref{eq:sig-ii-cal}).  The reader may easily verify that the unphysical IR
regularization parameter $\Delta$ cancels in the sum, as it should.

It is important to note that the angle $\vartheta_0'$ plays only the r\^{o}le
of an intermediate regulator for the bare event selection and drops out in the
final result.  In the calorimetric case there are no large logarithmic
contributions from final state radiation as long as $\vartheta_0'$ does not
become too small, since the mass singularity that is connected with the
emission of the photon off the scattered lepton is canceled in accordance
with the Kinoshita-Lee-Nauenberg theorem~\cite{KLN}.  For more details we
refer the reader to \cite{AAKM:nlo}.

%%%%%%%%%%%%%%%%%%%%%%%%%%%%%%%%%%%%%%%%%%%%%%%%%%%%%%%%%%%%%%%%%%%%%%%%

\section{Numerical Results}
\label{sec:results}

In order to illustrate our results, we shall now present some numerical values
obtained for the leading and next-to-leading radiative corrections.  To
facilitate the comparison of the results for the $\Sigma$ method with those
for determinations of the kinematic variables based on a lepton-only
measurement \cite{AAKM:ll,AAKM:nlo}, we used as input
\begin{equation}
  E_e = 27.5 \GeV \, , \quad
  E_p = 820  \GeV \, , \quad
  \vartheta_0 = 0.5 \; \mathrm{mrad} \, .
\end{equation}
Unless stated otherwise, we chose the ALLM97 parameterization \cite{ALLM97} as
structure function with $R=0$, no cuts were applied to the photon phase space,
and we assumed a calorimetric event selection.  For the sake of simplicity we
took a fixed representative angular resolution of $\vartheta_0' = 50 \;
\mathrm{mrad}$ for the electromagnetic calorimeter to separate nearby hits by
an electron or positron and a hard photon, which is close to realistic for the
H1 detector at HERA.  Also we disregard any effects due to the magnetic field
bending the scattered charged lepton away from a collinear photon.

\begin{figure}[p]
  \begin{center}
    \begin{picture}(120,120)
      % For PAW picture:
      \put(0,0){
        \includegraphics[width=120mm]{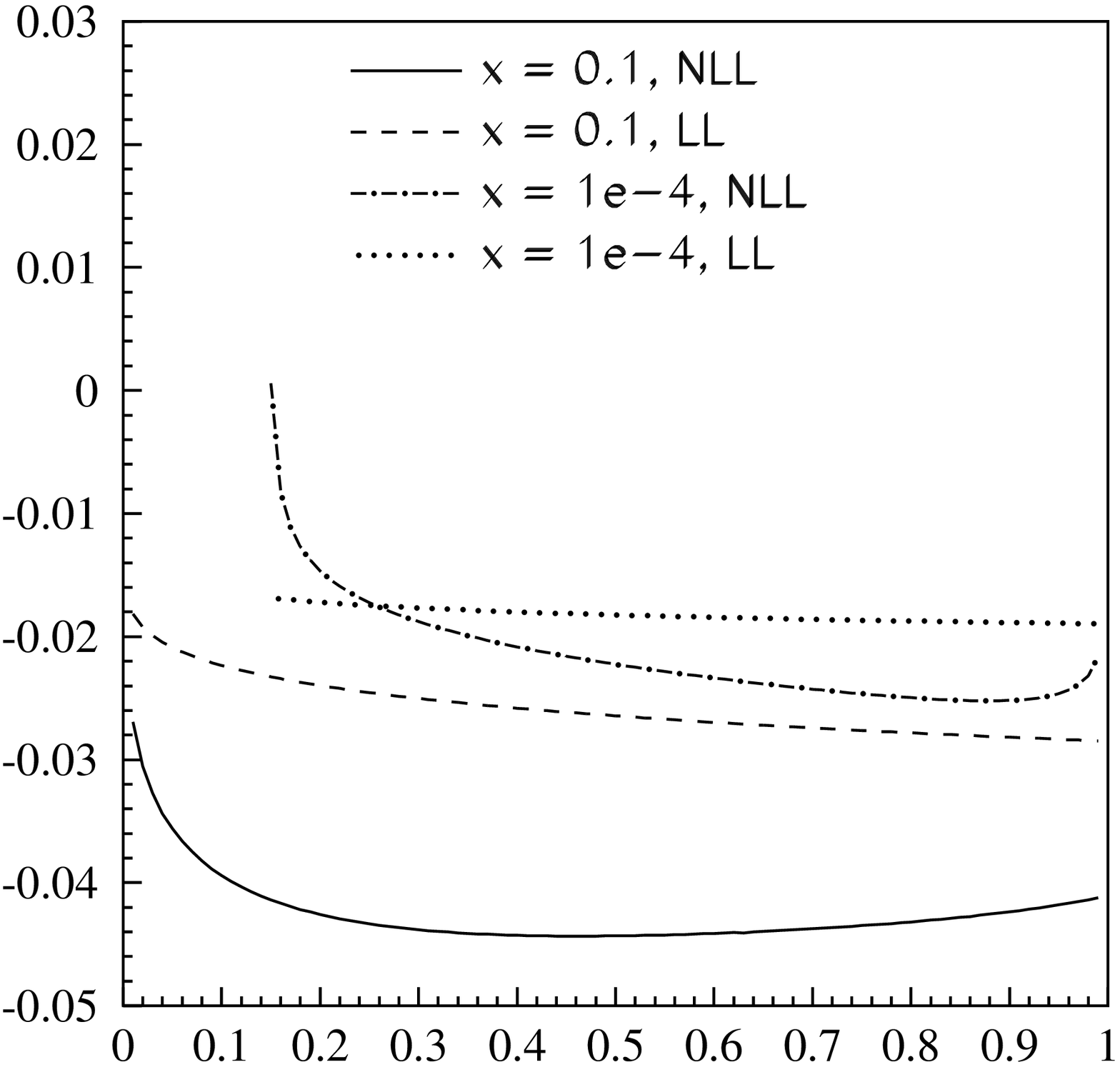}
        }
      \put(104,0){$\hat{y}$}
      \put(-6,101){$\delta_\mathrm{RC}$}
    \end{picture}
    \caption{Radiative corrections $\delta_\mathrm{RC}$
      (\protect\ref{eq:delta}) with leading and next-to-leading logarithmic
      accuracy at $\hat{x} = 0.1$ and $\hat{x} = 10^{-4}$ and a tagged photon
      energy of 5\GeV.  No cuts have been applied to the phase space of the
      second (semi-collinear) photon.}
    \label{fig:plot1}
  \end{center}
\end{figure}

\begin{figure}[p]
  \begin{center}
    \begin{picture}(120,120)
      % For PAW picture:
      \put(0,0){
        \includegraphics[width=120mm]{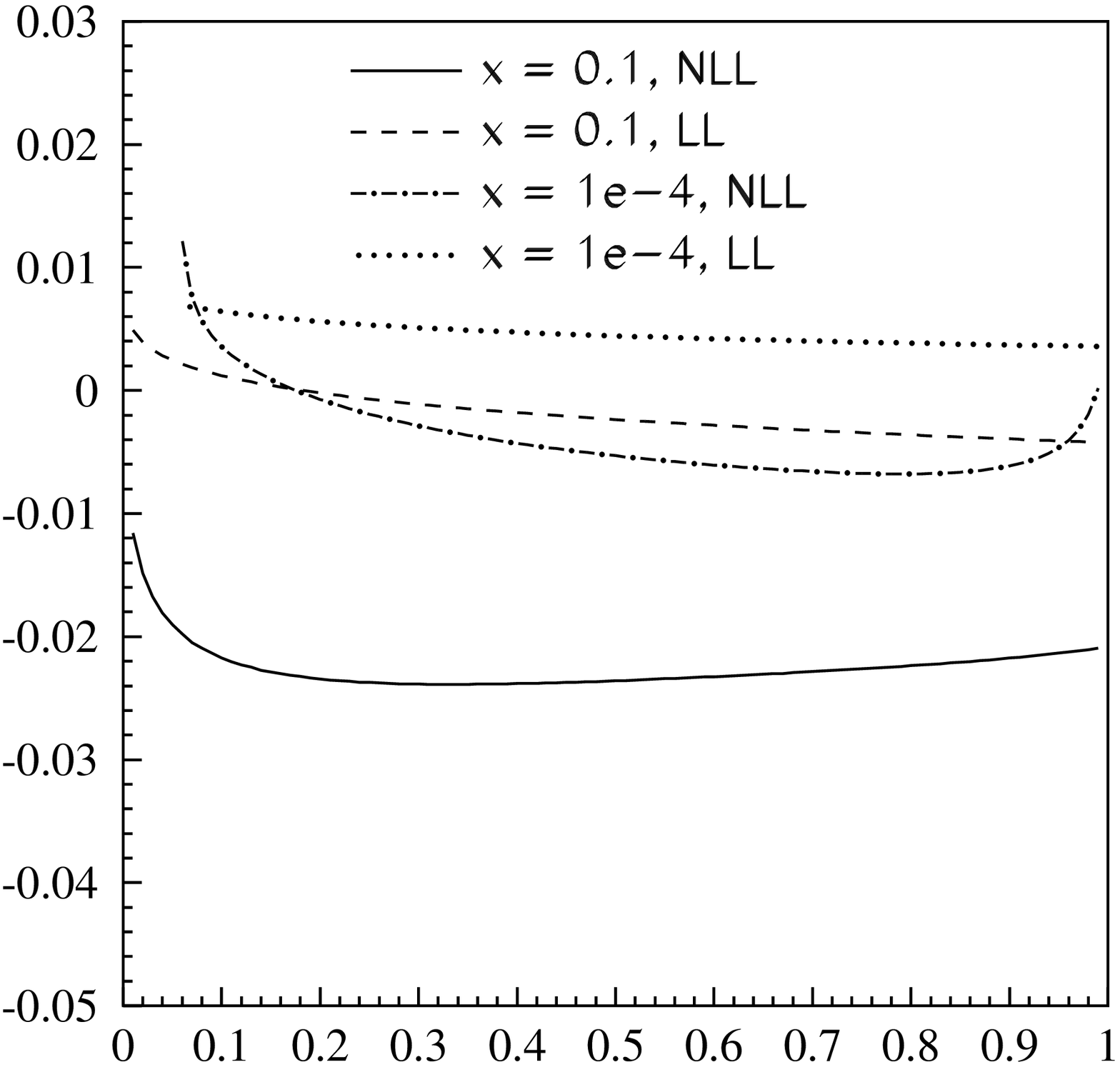}
        }
      \put(104,0){$\hat{y}$}
      \put(-6,101){$\delta_\mathrm{RC}$}
    \end{picture}
    \caption{Radiative corrections $\delta_\mathrm{RC}$
      (\protect\ref{eq:delta}) with leading and next-to-leading logarithmic
      accuracy at $\hat{x} = 0.1$ and $\hat{x} = 10^{-4}$ and a tagged photon
      energy of 20\GeV.  No cuts have been applied to the phase space of the
      semi-collinear photon.}
    \label{fig:plot2}
  \end{center}
\end{figure}

Figure~\ref{fig:plot1} compares the radiative correction
\begin{equation} \label{eq:delta}
  \delta_\mathrm{RC} =
  \frac{\dd^3\sigma}{\dd^3\sigma_\mathrm{Born}} - 1
\end{equation}
with leading and next-to-leading logarithmic accuracy at $\hat{x}=0.1$ and
$\hat{x}=10^{-4}$ and for a tagged energy of $E_\mathrm{PD} = 5\GeV$.  The
corresponding results for a tagged energy of $E_\mathrm{PD} = 20\GeV$ are
shown in figure~\ref{fig:plot2}.  The apparent cutoff at small $\hat{y}$ for
small $\hat{x}$ is due to touching of the narrow cones defined by the solid
angle covered by the photon detector and the cone around the final state
lepton.

At first sight the radiative corrections, being only of the order of a few
percent, look rather small as compared to, e.g., a leptonic measurement.
However, this apparent suppression of the leading logarithmic part of the QED
radiative corrections is easily traced back to the known weak dependence on
initial state radiation of the determination of the kinematic variables
(\ref{eq:kin-vars-measured})--(\ref{eq:kin-vars}) using the $\Sigma$ method
\cite{BB95,Wol97}.  On the other hand, the pure next-to-leading logarithmic
corrections are unsuppressed, as this ``cancellation mechanism'' does not work
for non-collinear photon radiation.

\begin{figure}[p]
  \begin{center}
    \begin{picture}(120,120)
      % For PAW picture:
      \put(0,0){
        \includegraphics[width=120mm]{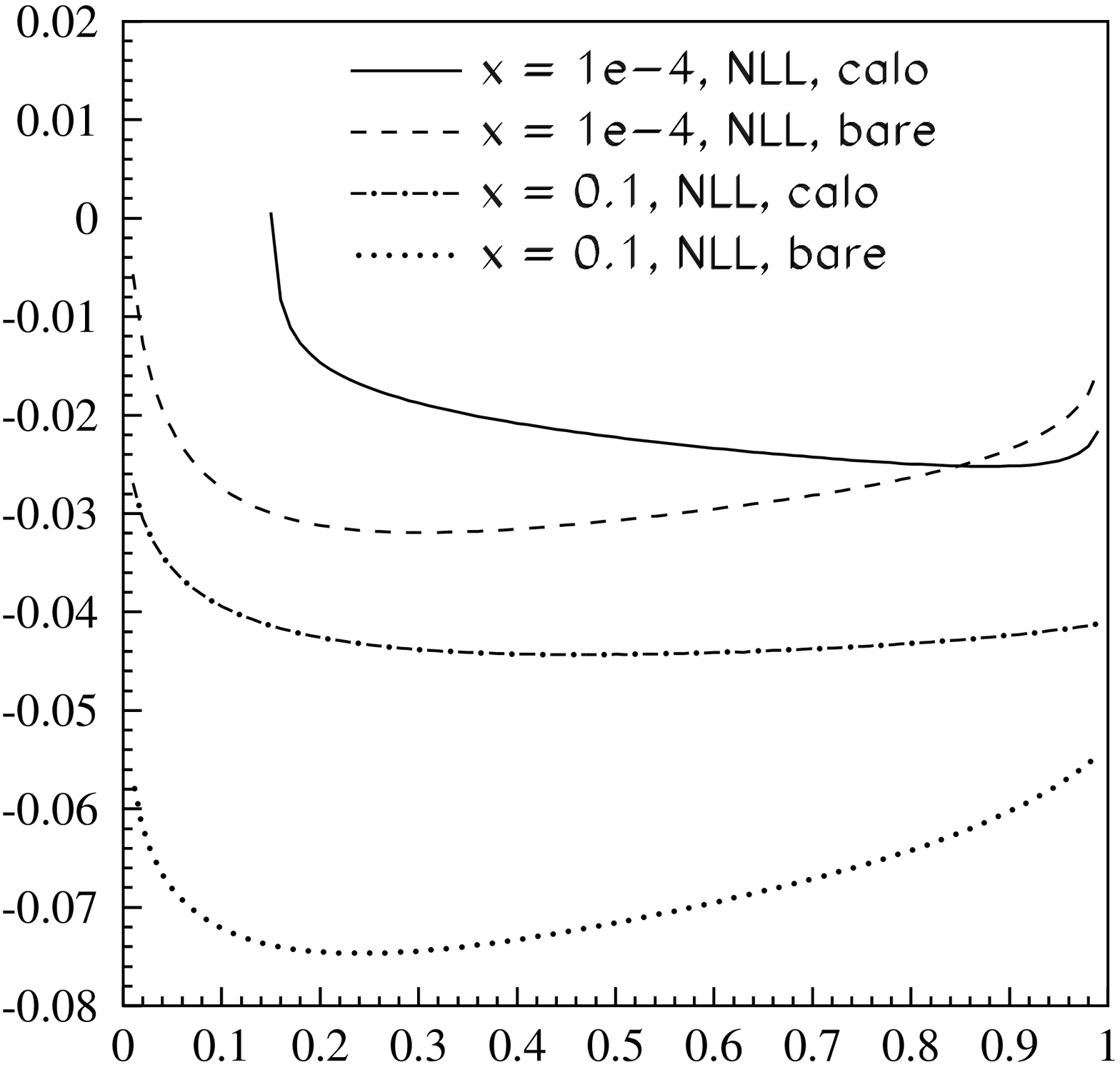}
        }
      \put(104,0){$\hat{y}$}
      \put(-6,103){$\delta_\mathrm{RC}$}
    \end{picture}
    \caption{Comparison of the radiative corrections for calorimetric
      (``calo'') vs.\ bare measurement of the scattered lepton at $\hat{x} =
      10^{-4}$ and at $\hat{x} = 0.1$ for a tagged photon energy of 5\GeV.}
    \label{fig:plot1a}
  \end{center}
\end{figure}

When we choose a bare event selection instead of a calorimetric one, the
radiative corrections do become slightly larger.  This is demonstrated in
figure~\ref{fig:plot1a} that compares the radiative corrections for both
selection schemes.  The difference between the curves may be easily understood
by noticing that in the calorimetric case the contribution from final state
corrections to the cross section is proportional to the relatively small
logarithm $\ln\vartheta'_0$, while it depends on the larger logarithm $\ln
\Eep/m$ in the bare case.

\begin{figure}[p]
  \begin{center}
    \begin{picture}(120,120)
      % For PAW picture:
      \put(0,0){
        \includegraphics[width=120mm]{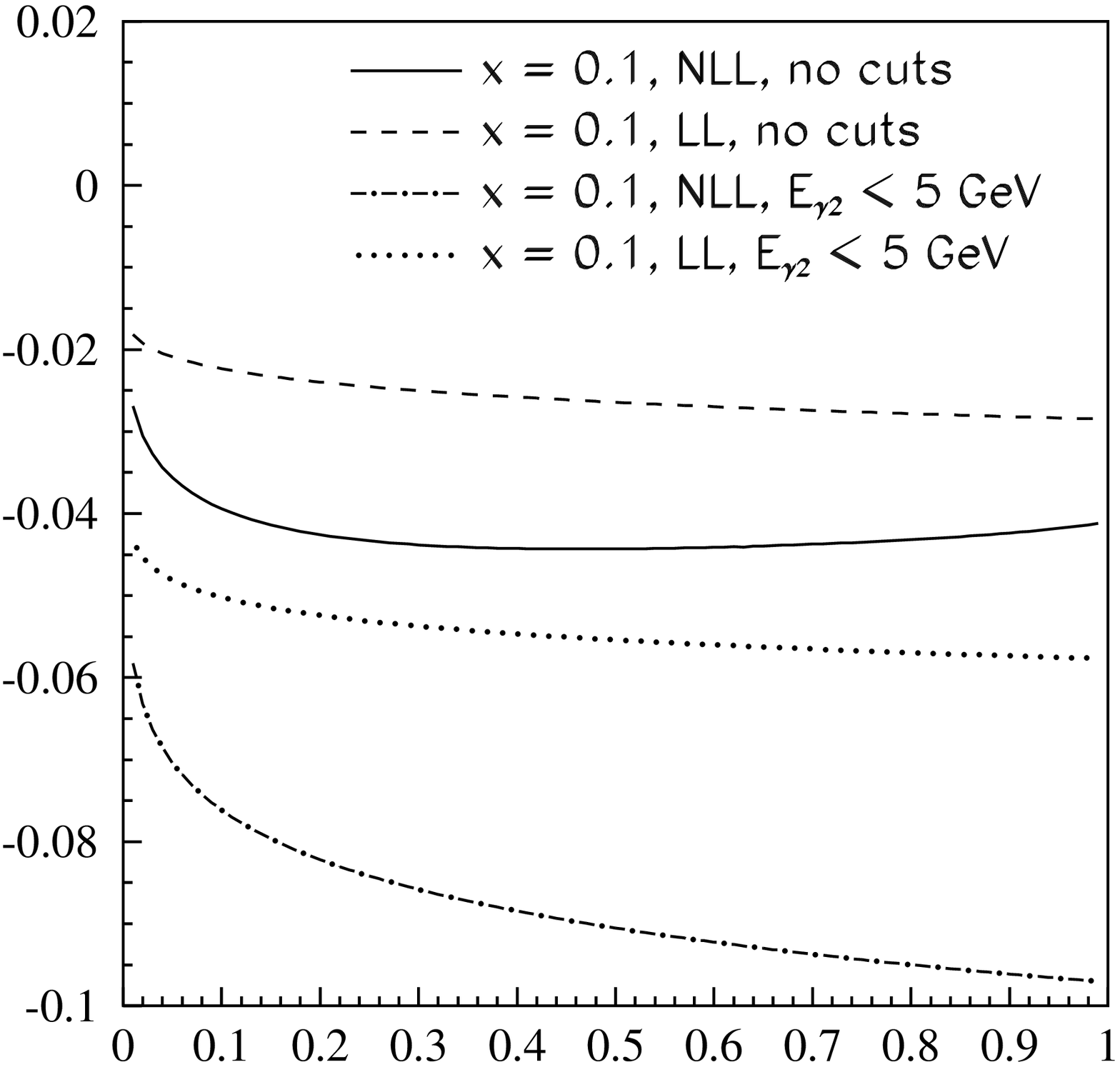}
        }
      \put(104,0){$\hat{y}$}
      \put(-6,101){$\delta_\mathrm{RC}$}
    \end{picture}
    \caption{Comparison of the cut dependence of the radiative corrections
      at $\hat{x} = 0.1$ for a tagged photon energy of 5\GeV.  A cut of
      $E_{\gamma2} < 5\GeV$ has been applied to the phase space of the
      semi-collinear photon.}
    \label{fig:plot3}
  \end{center}
\end{figure}

\begin{figure}[p]
  \begin{center}
    \begin{picture}(120,120)
      % For PAW picture:
      \put(0,0){
        \includegraphics[width=120mm]{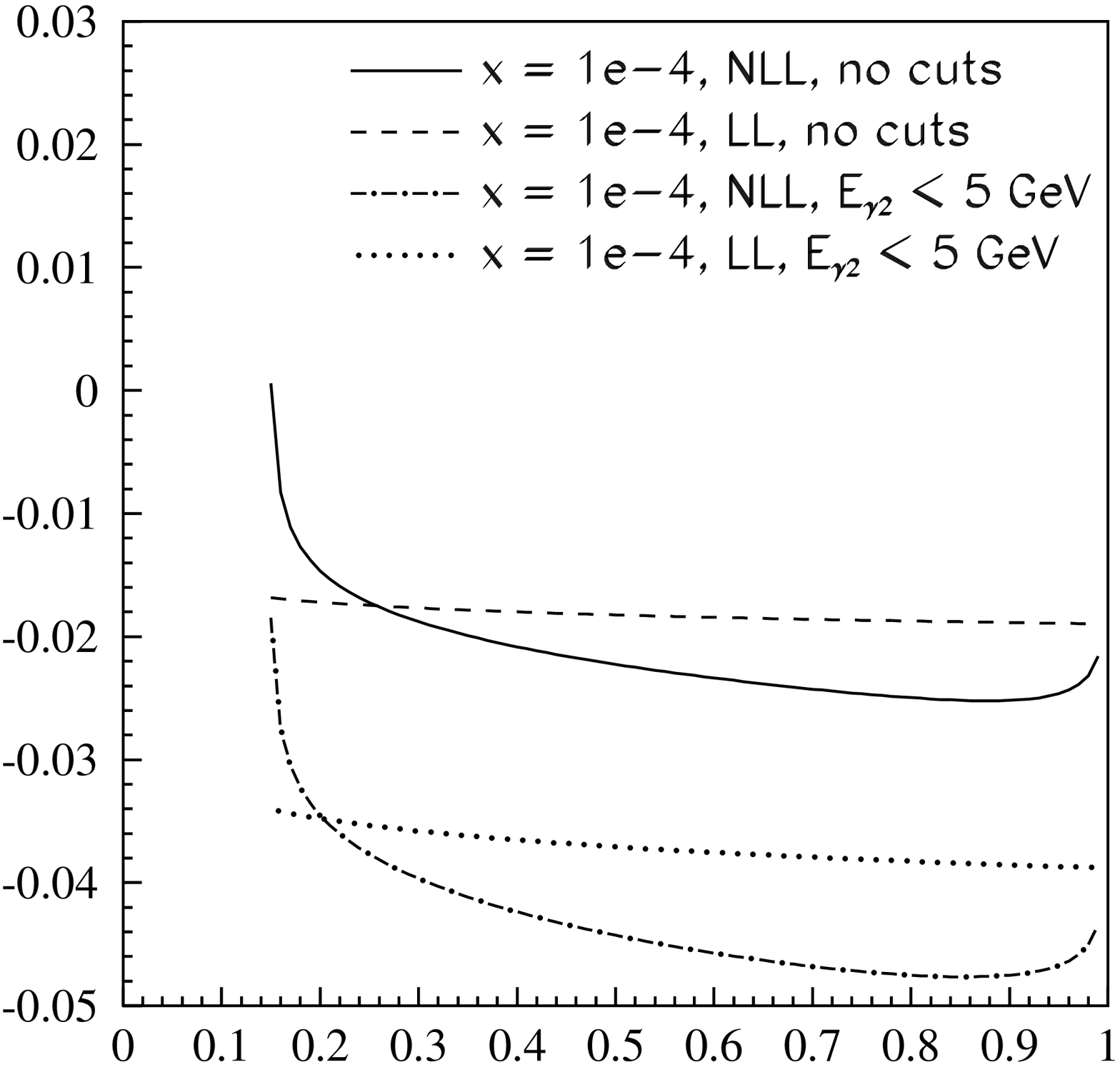}
        }
      \put(104,0){$\hat{y}$}
      \put(-6,101){$\delta_\mathrm{RC}$}
    \end{picture}
    \caption{Comparison of the cut dependence of the radiative corrections
      at $\hat{x} = 10^{-4}$ for a tagged photon energy of 5\GeV.  A cut of
      $E_{\gamma2} < 5\GeV$ has been applied to the phase space of the
      semi-collinear photon.}
    \label{fig:plot4}
  \end{center}
\end{figure}

Next we shall study the influence of a photon energy cut on the radiative
corrections.  For simplicity, we assume an emission angle independent cut
$E_{\gamma2}^\mathrm{max}$, which may be realized at large angles by rejecting
events that show energy in the electromagnetic calorimeter sufficiently
separated from the final state lepton, and for small angles by a cut on the
variable
\[
  \delta \, := \, \Sigma_h + \Eep (1+\cos\theta) - 2(E_e - E_\mathrm{PD}) \, ,
\]
which is about twice the energy of a photon that is lost outside the PD.
Figures~\ref{fig:plot3} and \ref{fig:plot4} illustrate the dependence of the
radiative corrections on an energy cut on the semi-collinear (lost) photon.
The influence of a rather loose cut of $E_{\gamma2} < 5\GeV$ is significant,
especially at the larger value of $\hat{x}=0.1$, although the inclusive
corrections were seen to be quite small.

\begin{figure}[p]
  \begin{center}
    \begin{picture}(120,120)
      % For PAW picture:
      \put(0,0){
        \includegraphics[width=120mm]{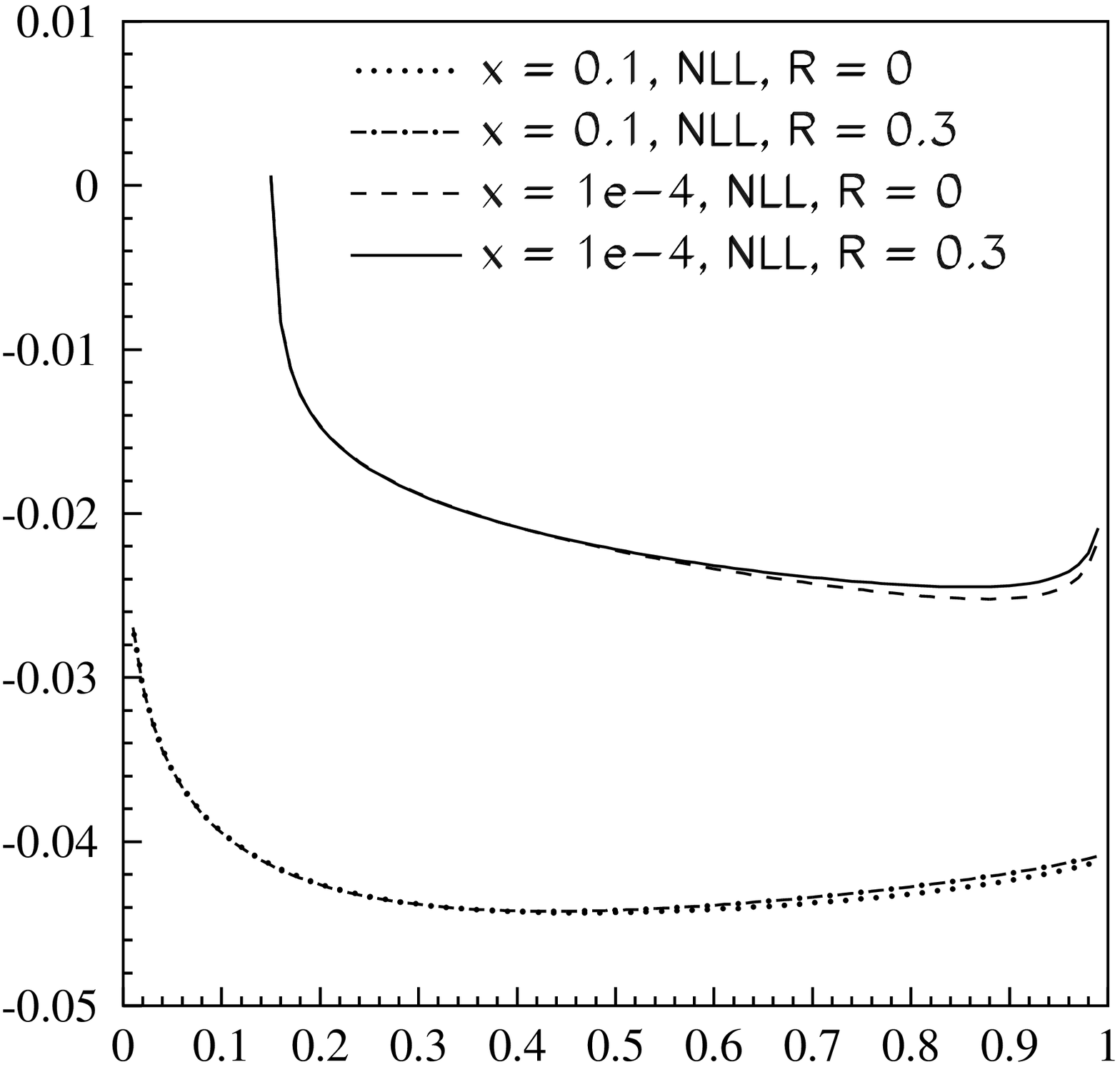}
        }
      \put(104,0){$\hat{y}$}
      \put(-6,101){$\delta_\mathrm{RC}$}
    \end{picture}
    \caption{Comparison of the $R$ dependence of the NLL corrections at
      $\hat{x} = 0.1$ and $\hat{x} = 10^{-4}$ for a tagged photon energy
      of 5\GeV.  No cut has been applied to the phase space of the
      semi-collinear photon.}
    \label{fig:plot5}
  \end{center}
\end{figure}

Finally, figure~\ref{fig:plot5} shows the dependence of the next-to-leading
logarithmic corrections on the ratio $R$ (see eq.~\ref{eq:R}), again for
$\hat{x}=0.1$ and $\hat{x}=10^{-4}$ and for a tagged energy of $E_\mathrm{PD}
= 5\GeV$.  As one would expect, one sees that only for large $\hat{y}$ there
is a visible difference between the corrections calculated for (assumed
constant) $R=0$ and $R=0.3$ of the order of a permille.  In the case of the
purely leading logarithmic corrections the change would be much less than the
order of the line width, so we omitted the corresponding lines.  For this
reason, a poorly known $R(x,Q^2)$ as input to the calculation of the QED
corrections will not have any significant effect on the extraction
\cite{FGMZ96} of $R$ from the measured tagged photon cross section.
Increasing the value of $R$ up to, say, $R=1$ for the smaller value of
$\hat{x}$ would simply increase the difference with respect to the curve for
$R=0$ but not lead to a qualitative change of our conclusion.

%%%%%%%%%%%%%%%%%%%%%%%%%%%%%%%%%%%%%%%%%%%%%%%%%%%%%%%%%%%%%%%%%%%%%%%%

\section{Summary}
\label{sec:sum}

The $\Sigma$ method for the determination of the kinematic variables in deep
inelastic scattering has been reviewed in some detail from a theoretical point
of view.  We derived the relevant kinematics for the calculation of the hard
photon emission contributions of the QED corrections to deep inelastic
scattering for the HERA collider.  As an application, we extended the
formalism to radiative DIS events with a hard photon tagged in the forward
photon detectors of the H1 and ZEUS experiments.  We have adapted the
calculations \cite{AAKM:ll,AAKM:JETP,AAKM:nlo} of the radiative corrections to
these DIS events with a tagged photon for the $\Sigma$ method determination of
kinematic variables.  It turned out that for a calorimetric measurement of the
final state lepton the leading-logarithmic corrections are suppressed and thus
quite small (of the order of 5\%), which is an intrinsic feature of the
$\Sigma$ method.  However, the typical size of the unsuppressed
next-to-leading logarithmic contributions is of the same order of magnitude.
However, for a bare event selection there are also significant contributions
to the corrections already at the leading logarithmic level.

The smallness of the leading QED corrections for the $\Sigma$ method suggests
that the corrections are well under control, so that neglected higher order
corrections will not play a significant r\^{o}le.  It is also encouraging that
the dependence of the corrections on the poorly known longitudinal structure
function $F_L$ (and thus $R$) is very small.

%%%%%%%%%%%%%%%%%%%%%%%%%%%%%%%%%%%%%%%%%%%%%%%%%%%%%%%%%%%%%%%%%%%%%%%%

\section*{Acknowledgments}

It is a great pleasure to thank A.B. Arbuzov, M. Fleischer and H. Spiesberger
for a critical reading of the manuscript and many useful suggestions.

%%%%%%%%%%%%%%%%%%%%%%%%%%%%%%%%%%%%%%%%%%%%%%%%%%%%%%%%%%%%%%%%%%%%%%%%

\appendix

\section{Kinematic Limit of Hard Photon Emission}
\label{app:kin-limit1}

This appendix is devoted to a discussion of the phase space limit for the
emission of a hard photon in radiative DIS in the case of the $\Sigma$ method.
As we shall refer exclusively to the $\Sigma$ method, and to avoid cumbersome
notation, we drop the index $\Sigma$ from all kinematic variables.

In the parameterization of the photon phase space following from
(\ref{eq:HERA-kin}), the invariant mass of the hadronic system (\ref{eq:W2})
is
\begin{eqnarray} \label{eq:W2lim1}
  W^2 & = &
        M^2
        + y S \left( 1 - \kappa - \frac{x}{1 - \kappa} \right)
        - 2 x_\gamma E_e^2      (1-\cos\vartheta)
        \nonumber \\
        && {} + 2 x_\gamma E_e \Eep (1-\cos\vartheta')
        \nonumber \\
        & = &
        M^2
        + y S \left[ 1 - \kappa - x \right]
        - 4 E_e^2 \, \frac{1-\cos\vartheta}{1+\cos\vartheta} \, \kappa
          \left[ 1 - (1 - \kappa)(1 - y)\right]
        \nonumber \\
        && {}
        - 4 \kappa E_e \sqrt{xy(1-y) S} \,
        \sqrt{\frac{1-\cos\vartheta}{1+\cos\vartheta}}
        \, \cos\varphi
        \, ,
\end{eqnarray}
where we heavily used the relations (\ref{eq:scprod2})--(\ref{eq:Eep}) and
\begin{equation} \label{eq:kap}
  \kappa = x_\gamma \, \frac{1 + \cos\vartheta}{2} \, .
\end{equation}
With the help of the following abbreviations,
\begin{eqnarray} \label{eq:abbrevs}
  \mu & = &
  1 - x - \frac{\bar{M}^2 - M^2}{yS} \equiv 1 - x - \frac{\bar\Delta_m}{y}
  \, , \nonumber \\
  \lambda & = &
  \sqrt{ \frac{E_e}{E_p} \frac{1-\cos\vartheta}{1+\cos\vartheta} }
  \, \nonumber \\
  \nu & = &
  \sqrt{ 4xy(1-y) } \, \cos\varphi
  \, ,
\end{eqnarray}
which satisfy the constraints
\begin{equation}
  0 \leq \mu < 1 \; , \qquad
  0 \leq \lambda < \infty \; , \qquad
  |\nu| < 1 \, ,
\end{equation}
the inequality $W^2 \geq \bar{M}^2$ can be brought into a simpler form:
\begin{equation} \label{ineq:kappa0-1}
  y \mu - \left[ y(1+\lambda^2) + \lambda\nu \right] \kappa
        - \lambda^2 (1-y) \kappa^2
        \geq 0  \, .
\end{equation}
Obviously the physical range is $0 \leq \kappa \leq \kappa_0$, with
\begin{equation} \label{eq:kappa0-1}
  \kappa_0 =
        - \frac{y(1+\lambda^2) + \lambda\nu}{2\lambda^2 (1-y)}
        + \frac{\sqrt{[y(1+\lambda^2) + \lambda\nu]^2 + 4y(1-y)\mu\lambda^2}}
        {2\lambda^2 (1-y)}
        \, .
\end{equation}
Direct inspection of (\ref{ineq:kappa0-1}) shows that for $y \to 1$,
$\kappa_0$ approaches the value
\begin{equation} \label{eq:tilde-kappa0}
  \tilde\kappa_0 \equiv \kappa_0(y=1) =
  \frac{1 - x - \bar\Delta_m}{1 + \lambda^2} < 1 \, .
\end{equation}

Note that the choice of $\vartheta$ and $\varphi$ for the parameterization of
the photon phase space is not well suited for analytic or semi-analytic
calculations, as the treatment of the separation between the phase space
regions \textit{ii)} and \textit{iii)} in section~3 is rather cumbersome.

%%%%%%%%%%%%%%%%%%%%%%%%%%%%%%%%%%%%%%%%%%%%%%%%%%%%%%%%%%%%%%%%%%%%%%%%

\section{Parameterization of the Hard Photon Phase Space}
\label{app:kin-limit2}

Instead of using the parameterization (\ref{eq:HERA-kin}) for the phase space
of the hard photon, it is often convenient to trade the azimuthal angle
$\varphi$ in favor of the angle $\vartheta'$ between the photon and the final
state lepton.  Introducing the variables $\tau_1, \tau_2$,
\begin{eqnarray} \label{def:taus}
  \tau_1 & := & \frac{1 - \cos\vartheta}{2} \, , \nonumber \\
  \tau_2 & := & \frac{1 - \cos\vartheta'}{2} =
  \frac{1 - \cos\theta \cos\vartheta - \sin\theta \sin\vartheta \cos\varphi}{2} \, ,
\end{eqnarray}
the integration over the photon solid angle becomes
\begin{equation} \label{eq:dOmega}
  \int\!\! \dd\Omega_\gamma \equiv
  \int\!\! \dd(\cos\vartheta) \, \dd\varphi =
  \int\!\! \mathcal{J}(\tau_1,\tau_2) \, \Theta(\mathcal{D}) \,
  \dd\tau_1 \, \dd\tau_2 \, ,
\end{equation}
with the Jacobian
\begin{eqnarray}
  \mathcal{J}(\tau_1,\tau_2) & = &
  \frac{8}{|\sin\theta \sin\vartheta \sin\varphi |} =:
  \frac{4}{\sqrt{\mathcal{D}}} \, ,
  \nonumber \\
  \noalign{\hbox{where}}
  \label{eq:D}
  \mathcal{D} & = &
  - 4 \tau \tau_1 \tau_2
  - \left(
    \tau^2 + \tau_1^2 + \tau_2^2
    - 2 \tau \tau_1 - 2\tau \tau_2 - 2\tau_1 \tau_2
  \right) ,
  \\
  \label{eq:tau}
  \tau & := & \frac{1 - \cos\theta}{2}
  \; = \; \frac{x y \vareps}{(1-\kappa)^2 (1 - y) + x y \vareps}
  \, , \\
  \vareps & := & \frac{E_p}{E_e} \, . \nonumber
\end{eqnarray}
A factor of 2 has been taken into account in the Jacobian for the two-fold
ambiguity in the azimuthal angle $\varphi$.  Note that $0\leq\kappa<1$ implies
\begin{equation} \label{eq:tau0}
  \tau \geq
  \tau^{(0)}(x,y) \, \equiv \,
  \frac{x y \vareps}{1 - y + x y \vareps} \, .
\end{equation}
The range of integration for the variables $\tau_{1,2}$ for a given photon
energy $x_\gamma E_e$ follows from the argument of the step function
$\Theta(\mathcal{D})$.  It is trivially obtained if $\tau$ depends only on the kinematic
variables $x,y$ but not on the photon phase space variables, as is the case
for a purely leptonic measurement of the kinematic variables \cite{AAKM:nlo},
but it is more involved for the $\Sigma$ method and will therefore be
discussed below.

%%% Discussion of the mathematical consequences by the change of variables

As the relevant variables for the parameterization of the phase space of the
photon we choose $\tau_1$, $\tau_2$, and $x_\gamma$.  Due to (\ref{eq:tau}),
$\tau$ is related to $x_\gamma$, as $\kappa = x_\gamma(1-\tau_1)$.  Therefore,
it appears to be reasonable to take $\tau_1$ as the outermost integration and
to determine the range of integration for the other two variables for each
value of $\tau_1$, $0<\tau_1<1$.

\begin{figure}[p]
  \begin{center}
    \vspace*{10mm}
    \includegraphics[width=80mm]{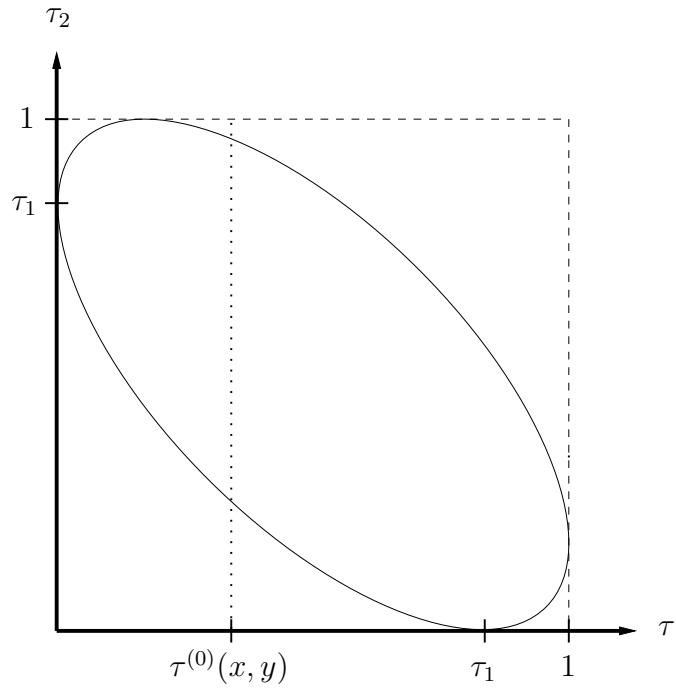}
    \begin{picture}(80,0)(0,0)
      \put(82,6.5){$\tau$}
      \put(57,1){$\tau_1$}
      \put(69,1){$1$}
      \put(-4,63){$\tau_1$}
      \put(-3,74){$1$}
      \put(0.5,88){$\tau_2$}
      \put(17,1){$\tau^{(0)}(x,y)$}
    \end{picture}
    \caption{Schematic view of the kinematic range for the integration over
      the angles $\tau$ and $\tau_2$ for fixed $\tau_1$.  The allowed range
      from the change of variables (\ref{def:taus}) alone is given by the
      interior of the ellipse.  In addition, for given $x,y$, the region left
      to the dotted line is unphysical, as $\tau \geq \tau^{(0)}(x,y)$ from
      (\ref{eq:tau0}).}
    \label{fig:ellipse}
  \end{center}
\end{figure}

Obviously, the range where the argument $\mathcal{D}$ (\ref{eq:D}) of the
$\Theta$ function above is positive is given by the interior of an ellipse as
shown in figure~\ref{fig:ellipse}.  The ellipse touches the $\tau$- and
$\tau_2$-axes at the value of $\tau_1$, and the lines $\tau=1$ and $\tau_2=1$
at the value of $(1-\tau_1)$.

The explicit form of the upper and lower boundaries of the ellipse are
\begin{equation} \label{eq:taupm}
  \tau_2^\pm(\tau_1,\tau) \, = \,
  \tau_1 (1-\tau) + \tau(1-\tau_1)
  \pm 2 \sqrt{\tau_1 (1-\tau) \tau (1-\tau_1)} \, .
\end{equation}

On the other hand, remembering that for given $x,y$ there is a lower limit
(\ref{eq:tau0}) on $\tau$, we see that the part of this ellipse left to the
dotted line is certainly unphysical.

%%%

Yet we have not made use of a lower limit on the hadronic mass that is due to
the inelastic threshold or to a lower cut on the invariant hadronic mass.  To
this end, we express the hadronic mass (\ref{eq:W2lim1}) in terms of the
$\tau_{1,2}$:
\begin{eqnarray} \label{eq:W2lim2}
  W^2 & = &
  M^2
        + y S \left( 1 - \kappa - \frac{x}{1 - \kappa} \right)
        - 4 x_\gamma E_e^2 \tau_1
        + 4 x_\gamma E_e \Eep \tau_2
        \nonumber \\
        & = & M^2
        + y S \left[ 1 - \kappa - \frac{x}{1 - \kappa} \right]
        - \frac{S}{\vareps} \,
        \frac{\tau_1}{1 - \tau_1}
        \, \kappa
        \\
        && {} + S \,
        \frac{\tau_2}{1 - \tau_1}
        \left[
        \frac{(1-\kappa)(1-y)}{\vareps} + \frac{xy}{1-\kappa}
        \right] \kappa \nonumber \; .
\end{eqnarray}
Requiring $W^2 \geq \bar{M}^2$ leads to the inequality
\begin{eqnarray} \label{ineq:W2lim}
  W^2 - \bar{M}^2
        & = &
        \frac{S}{(1-\kappa)(1-\tau_1)}
        \left[
        A \kappa^3 + B \kappa^2 + C \kappa + D
        \right]
        \geq 0
        \nonumber \\
\noalign{\hbox{where}}
        A & = & \frac{(1-y) \tau_2}{\vareps}
        \nonumber \\
        B & = & (1-\tau_1) y + \frac{\tau_1 - 2(1-y) \tau_2}{\vareps}
        \\
        C & = & xy \tau_2 - 2y(1-\tau_1) + (1-\tau_1) \bar\Delta_m
                + \frac{(1-y) \tau_2 - \tau_1}{\vareps}
        \nonumber \\
        D & = & (1-\tau_1) \left[ (1-x) y - \bar\Delta_m \right]
        \; . \nonumber
\end{eqnarray}
Since $A \geq 0$, $D>0$, the cubic equation $A \kappa^3 + B \kappa^2 + C
\kappa + D = 0$ will have either one or three real solutions: always a
negative one which is obviously unphysical, and two positive or complex
conjugate ones.

In case there are three real solutions, only the smaller one of the two
positive solutions may be physical.  This may be seen by studying the limit
$\tau_1 \to 1$, where $\kappa$ has to go to 0 for any finite photon energy
$x_\gamma$.

For small values of $\tau_2$, i.e., radiation collinear to the outgoing
lepton, one always has three real solutions.  This is easily verified by
direct inspection of (\ref{ineq:W2lim}) for $A \to 0$: the negative solution
goes to $-\infty$ essentially as $-B/A$, and the resulting quadratic equation
in the limit $\tau_2=0$ always has a positive discriminant.

In the remaining cases when there is only one real (and negative) solution,
the effective upper limit on $\kappa$ may be indirectly obtained from
(\ref{eq:tau}) and the maximum allowed value for $\tau^+$ as a function of
$\tau_{1,2}$, similarly to (\ref{eq:taupm}),
\begin{equation} \label{eq:limkappa}
  \kappa \, \leq \,
  \bar{\kappa}(\tau_1,\tau_2) \, \equiv \,
  1 -
  \sqrt{ \frac{1-\tau^+}{\tau^+}
         \frac{\tau^{(0)}}{1-\tau^{(0)}} } \, ,
\end{equation}
which is entirely of geometric origin.

Let us now compare the above limits that were derived from the lower limit (or
a cut) on the hadronic mass with the kinematic limits for collinear photon
radiation, as given in \cite{HECTOR}.
The case of collinear initial state radiation (ISR) is recovered by setting
$\tau_1 \to 0$, so that $\kappa \to x_\gamma$, $\tau_2 \Eep \to xy E_p /
(1-x_\gamma)$, and from (\ref{eq:W2lim2}) one immediately obtains
\begin{equation}
  x_\gamma \; \leq \;
        1 - x - \frac{\bar\Delta_m}{y}
        \, .
\end{equation}
This is consistent with the limit on $z$ ($= 1 - x_\gamma$) given in
\cite{HECTOR} up to terms of order $\mathcal{O} \left(\bar\Delta_m\right)$, as
these authors did not consider neither the inelastic threshold nor a cut on
the hadronic mass.

Collinear final state radiation (FSR) corresponds to taking the limit $\tau_2
\to 0$, $\tau_1 \to \tau = xy\vareps / \left\{ [1-x_\gamma(1-\tau)]^2(1-y) +
  xy\vareps \right\}$, with $\tau$ being implicitly defined.  However, in
order to be able to compare our limit with the one given in
ref.~\cite{HECTOR}, we shall think of the collinear photon as taking away the
fraction $(1-\zeta)$ of the outgoing lepton-photon system, while the lepton
retains the fraction $\zeta$.  Hence, we have $k = [(1-\zeta)/\zeta] p'$, and
we may write
\begin{equation}
  P \cdot k = \frac{1-\zeta}{\zeta} \, P \cdot p'
  \quad \Longrightarrow \quad
  \frac{\kappa}{1-\kappa}\frac{\zeta}{1-\zeta} = 1 - y
  \, , \quad
  1 - \kappa = \frac{\zeta}{1 - y(1-\zeta)}
  \, ,
\end{equation}
and thus
\begin{equation}
        p \cdot k =
        \frac{1-\zeta}{\zeta} \, p \cdot p' =
        \frac{x y}{1-\kappa} \frac{1-\zeta}{\zeta}   \cdot \frac{1}{2} S =
        \frac{x y [1-y(1-\zeta)](1-\zeta)}{\zeta^2} \cdot \frac{1}{2} S \, .
\end{equation}
Inserting these relations into (\ref{eq:W2}) we obtain
\begin{equation} \label{eq:lim-fsr}
  W^2 \Bigr|_\mathrm{FSR}
        \; = \; M^2 +
        \frac{\zeta^3 - x [ 1-y(1-\zeta)]^2}{\zeta^2[1 - y(1-\zeta)]} \, yS
        \; \geq \; \bar{M}^2
        \, .
\end{equation}
This inequality leads to a lower limit $\zeta_0$, $0 < \zeta_0 < 1$, that is
the single real solution of a cubic equation, $W^2(\zeta_0) = \bar{M}^2$.  The
kinematic limit that follows from (\ref{eq:lim-fsr}) is consistent with the
limit for FSR given in \cite{HECTOR}, again up to terms of order $\mathcal{O}
\left(\bar\Delta_m\right)$ that have been neglected in their work.

Finally, we should mention that the above considerations directly apply only
to the case of conventional deep inelastic scattering with a single radiated
photon.  However, the case of DIS with a tagged collinear photon and a hard
non-collinear second photon is easily recovered, provided we perform the
simultaneous substitutions
\begin{equation} \label{eq:subst}
  E_e \to z E_e
  \, , \quad
  S \to zS
  \, , \quad
  x_\gamma \to \frac{x_2}{z}
  \, , \quad
  \kappa \to \frac{\kappa}{z}
  \, , \quad
  \vareps \to \frac{\vareps}{z}
  \, , \quad
  \mbox{and} \quad
  \bar\Delta_m \to \frac{\bar\Delta_m}{z}
\end{equation}
in the expressions given above.

%%%%%%%%%%%%%%%%%%%%%%%%%%%%%%%%%%%%%%%%%%%%%%%%%%%%%%%%%%%%%%%%%%%%%%%%

%%%%%%%%%%%%%%%%%%%%%%%%%%%%%%%%%%%%%%%%%%%%%%%%%%%%%%%%%%%%%%%%%%%%%%%%

%\listoffigures

%%%%%%%%%%%%%%%%%%%%%%%%%%%%%%%%%%%%%%%%%%%%%%%%%%%%%%%%%%%%%%%%%%%%%%%%

%%%%%%%%%%%%%%%%%%%%%%%%%%%%%%%%%%%%%%%%%%%%%%%%%%%%%%%%%%%%%%%%%%%%%%%%

%%%%%%%%%%%%%%%%%%%%%%%%%%%%%%%%%%%%%%%%%%%%%%%%%%%%%%%%%%%%%%%%%%%%%%%%

%%%%%%%%%%%%%%%%%%%%%%%%%%%%%%%%%%%%%%%%%%%%%%%%%%%%%%%%%%%%%%%%%%%%%%%%

%%%%%%%%%%%%%%%%%%%%%%%%%%%%%%%%%%%%%%%%%%%%%%%%%%%%%%%%%%%%%%%%%%%%%%%%

%%%%%%%%%%%%%%%%%%%%%%%%%%%%%%%%%%%%%%%%%%%%%%%%%%%%%%%%%%%%%%%%%%%%%%%%

%%%%%%%%%%%%%%%%%%%%%%%%%%%%%%%%%%%%%%%%%%%%%%%%%%%%%%%%%%%%%%%%%%%%%%%%

\end{document}